\documentclass[12pt,preprint]{aastex62}
\usepackage{CJK}
\usepackage{xcolor}
\usepackage{multirow}
\begin{document}
\begin{CJK*}{UTF8}{gbsn}
\title{Evolution of the Quiescent Disk surrounding a Superoutburst of the Dwarf Nova TW\,Virginis}
\author[0000-0002-4280-6630]{Zhibin Dai (戴智斌)}
\affiliation{Yunnan Observatories, Chinese Academy of Sciences, 396 Yangfangwang, Guandu District, Kunming, 650216, P. R. China.}
\affiliation{Key Laboratory for the Structure and Evolution of Celestial Objects, Chinese Academy of Sciences, 396 Yangfangwang, Guandu District, Kunming, 650216, P. R. China.}
\affiliation{Center for Astronomical Mega-Science, Chinese Academy of Sciences, 20A Datun Road, Chaoyang District, Beijing, 100012, P. R. China.}
\affiliation{University of Chinese Academy of Sciences, No.19(A) Yuquan Road, Shijingshan District, Beijing, 100049, P.R.China}
\author[0000-0003-4373-7777]{Paula Szkody}
\affiliation{University of Washington, Seattle, WA, 98195, USA.}
\author[0000-0003-4069-2817]{Peter M. Garnavich}
\affiliation{University of Notre Dame, Notre Dame, IN, 46556, USA.}
\correspondingauthor{Zhibin Dai}
\email{zhibin\_dai@ynao.ac.cn}

\begin{abstract}
Portions of the Kepler K2 Short Cadence light curve of the dwarf nova (DN) TW\,Vir at quiescence are investigated using light curve modeling. The light curve was separated into 24 sections, each with a data length of $\sim\,$0.93\,d, comprising 4 sections before and 20 after a superoutburst (SO). Due to the morphological differences, the quiescent orbital modulation is classified into three types. Using a fixed disk radius and the two component stellar parameters, all 24 synthetic disk models from the sections show a consistent configuration, consisting of a disk and two hotspots: one at the vertical side of the edge of the disk and the other one on the surface of the disk. Before the SO, the disk and a ringlike surface-hotspot are suddenly enhanced, triggering a precursor and then SO. At the end of the quiescent period following the SO and before the first normal outburst, the edge-hotspot becomes hotter, while the surface-hotspot switches into a ``coolspot" with a coverage of nearly one-half of the disk surface. During quiescence, the surface-hotspot is always located at the outer part of the disk with a constant radial width. A flat radial temperature distribution of the disk is found and appears flatter when approaching the outburst. Like many U\,Gem-type DN with orbital periods of 3-5\,hr, the mass transfer rate is significantly lower than the predictions of the standard/revised models of CV evolution.
\end{abstract}

\section{Introduction}
\label{sec:sec1}

Dwarf novae (DN) constitute a subclass of cataclysmic variables (CVs; \citet{war03}), semi-detached interacting binaries in which a Roche-lobe filling secondary star transfers matter to a more massive white dwarf with a weak magnetic field (B$<$$10^{6}$\,G). The transferred material spreads into a disk surrounding the white dwarf via a viscous process and a hot and bright spot is created where the mass transfer stream strikes the disk. One interesting observational feature of DN is the existence of two types of large-amplitude and quasi-periodic outbursts: frequent normal outbursts (NO) and occasional superoutbursts (SO). Compared with a NO, the SO has a larger amplitude ($\sim$0.7\,mag higher than the maximal light of a NO) and longer duration time (about two weeks). 

Based on a typical $\alpha$-disk model \citep{sha73}, both the NO and SO are described as viscosity changes. The disk instability model \citep[DI;][]{osa74,las01} and the mass-transfer instability model \citep[MTI;][]{bat75} have been used to describe the NO. The thermal-tidal instability model \citep[TTI;][]{osa89,osa96}, the enhanced mass-transfer model \citep[EMT;][]{sma04,sma08,sma09} and the pure thermal instability model \citep[PTI;][]{can10,can12} were developed to explain the SO. Many tests of these models \citep[see,][]{sma13,osa13,osa14} have been done to judge the most plausible mechanism for the NO and SO \citep{ida10,ram17,dub18}. Most of the tests support the DI model, that a NO with an amplitude of 2$\sim$5\,mag lasting a few days to a few weeks, is attributed to the continuous pileup of the disk material transferred from the secondary star, and when an annulus of disk exceeds the local critical surface density, the propagation of a heating front rapidly ignites the whole disk. This means that the NO switches from a cold and low-viscosity disk to a hot and high-viscosity disk. Further modifications to the DI model are required to solve the remaining problems \citep[cf.][]{sma00,bak14}.

TW\,Vir is a well known U\,Gem-type DN \citep{oco32}. The short cadence (SC; 1 minute sampled) light curve observed during Campaign 1 of the Kepler K2 mission (K2-C1; \citet{how14}) observed a complete SO and two incomplete NO \citep{dai16}. \citet{dai17} used the quiescent portions of this light curve to improve the orbital period of TW\,Vir to 0.182682(3)\,d, and search for orbital modulation and found a double-hump orbital modulation similar to many other low-inclination DN (e.g. 1RXS\,J0632+2536 and RZ\,Leo \citep{dai16}). Using a synthetic program XRBinary developed by E. L. Robinson\footnote{\url{http://www.as.utexas.edu/~elr/Robinson/XRbinary.pdf}} and a nonlinear fitting code NMfit, \citet{dai18} were able to reproduce the averaged double-hump orbital modulation of TW\,Vir with a disk model consisting of a disk and two hotspots (one at the vertical side of the edge of the disk (hereafter, edge-hotspot\footnote{The impact region caused by the stream of gas from the secondary star.}), and the other one on the surface of the disk (hereafter, surface-hotspot\footnote{The surface-hotspot may arise from several physical mechanisms: the gas stream overflowing the disk edge and falling onto the top of the disk, spiral patterns or shocks on the disk surface, magnetic filed reconnections and so on.})). Although XRBinary was initially used to model the light in X-ray binaries, \citet{dai18} adapted this program to include low-inclination CV systems, where the unobscured white dwarf can be approximated by an unresolved, black-body radiator. In the best-fit disk model of TW\,Vir \citep{dai18}, the size of disk is $\sim$\,5600 times larger than that of the white dwarf. The scale of the white dwarf is close to one piece of the disk surface title since DISKTILES=20000. For TW\,Vir, the relative flux contribution from the white dwarf is $<$\,0.037 far smaller than those from the surrounding disk and the companion star. Thus, this small and faint white dwarf can be appropriately assumed to be an irradiating point-source. The best-fit disk model indicated that TW\,Vir is a low-inclination (44.3$^{\circ}$) DN with a white dwarf mass of 1.10\,M$_{\odot}$ and a red dwarf mass of 0.45\,M$_{\odot}$, consistent with previous results derived from the infrared and UV data \citep{cor82,mat85}. In this model, the disk contributes the largest luminosity ($>$63\% of the system light).

\citet{dai18} also determined that the orbital modulation was not stable. The smooth orbital modulation derived by \citet{dai17} did not repeat during the $\sim$20\,d after the SO, which implies that the disk does not accumulate material smoothly during quiescence. The general disk model derived from the averaged double-hump light curve \citep{dai18} did not explain the variations in the quiescent disk. This paper focuses on a detailed modeling effort of the short timescale disk and hotspot changes around a SO using the continuous high-precision K2-C1 data. Section~\ref{sec:sec2} introduces the phased and binned K2-C1 SC light curves and their classifications based on the light curve morphology. Sections~\ref{sec:sec3} describes the disk models obtained and a complete evolution scenario of the quiescent disk surrounding the SO is detailed in Section~\ref{sec:sec4}. The hotspots before the NO, disk temperature distribution and mass transfer rate at quiescence are also discussed.

\section{Light curve morphology}
\label{sec:sec2}

\begin{figure}
\centering
\includegraphics[width=16.0cm]{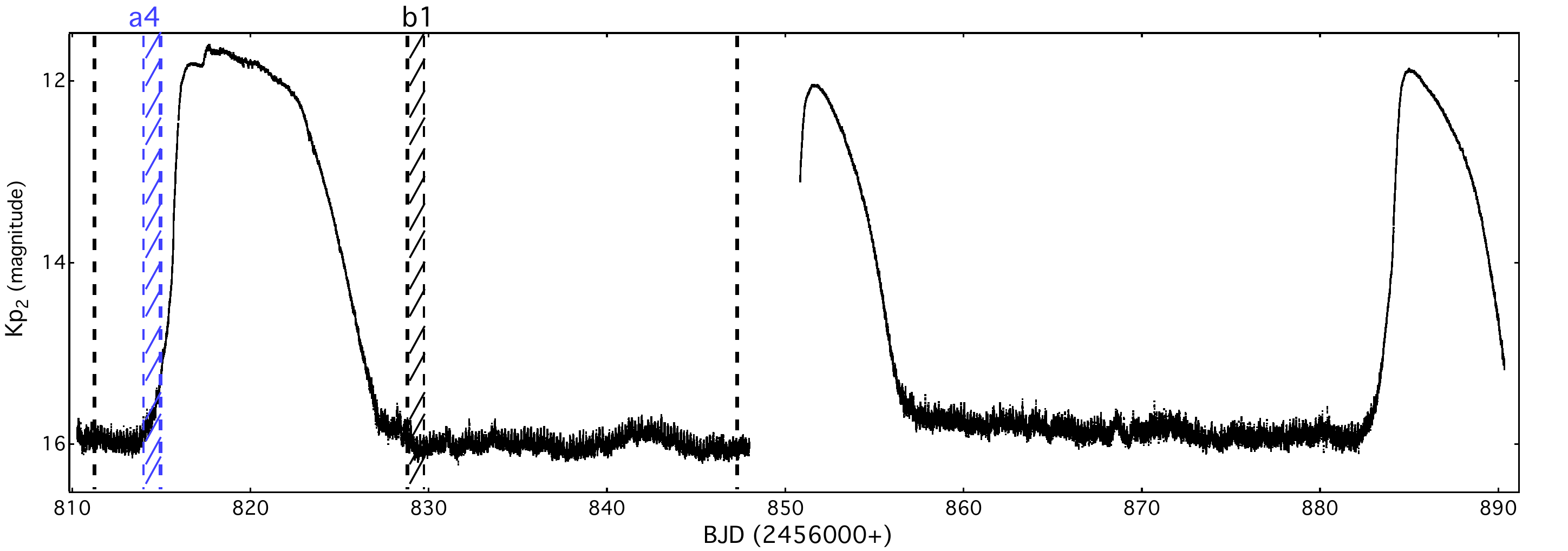}
\caption{The complete K2-C1 SC light curve of TW\,Vir in K2 magnitudes. The stable quiescent light curves before the first NO covering a duration of $\sim$\,35\,d interrupted by the SO lasting $\sim$\,15\,d are divided into two parts: pre-SO and post-SO indicated by the bold vertical dashed lines. The blue and black dashed areas refer to the last section (a4) before the SO and first section (b1) after the SO, respectively.}
\label{lc1}
\end{figure}

\begin{figure}
\centering
\includegraphics[width=14.0cm]{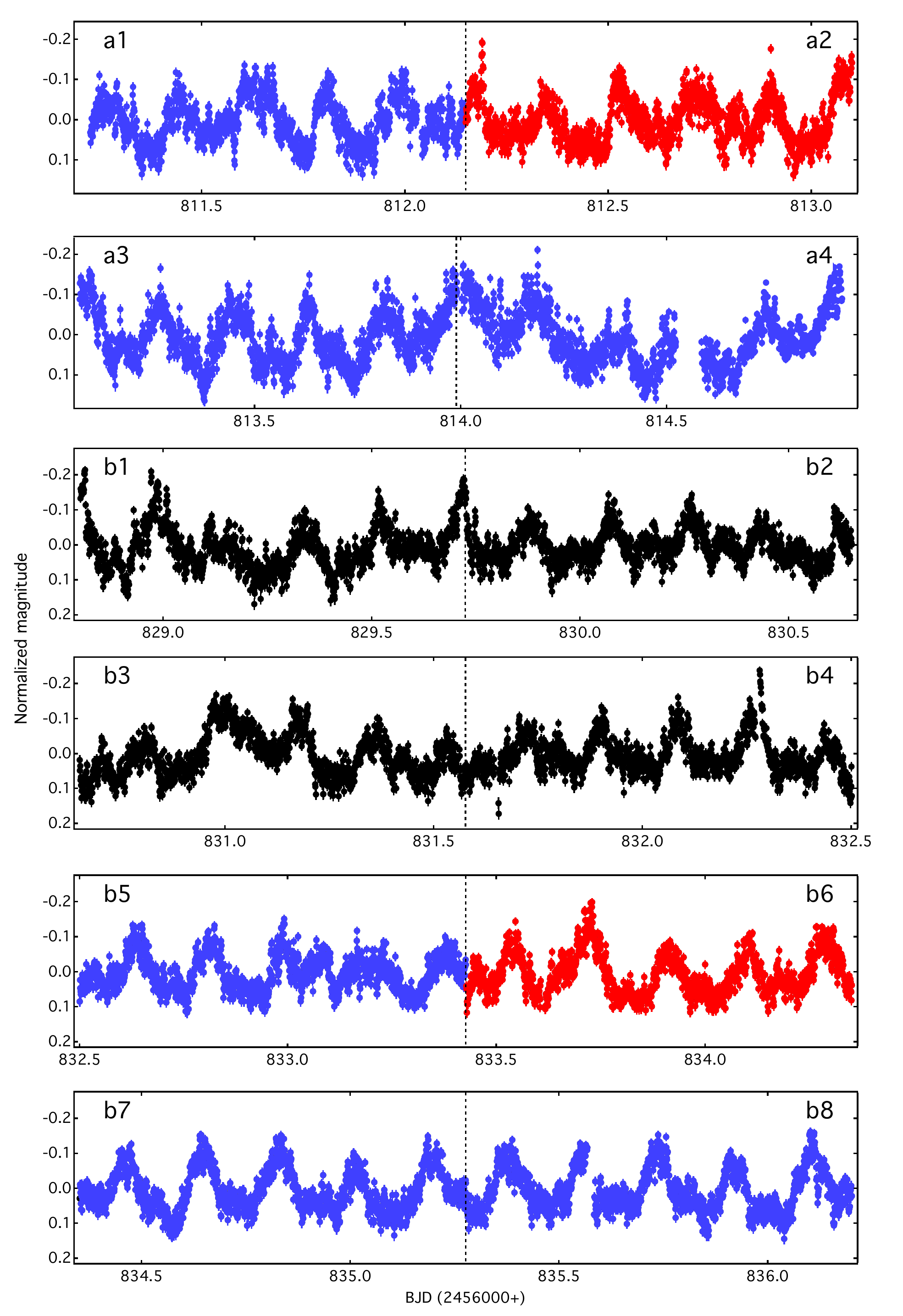}
\caption{The first half of the quiescent 24 sections (a1-a4 and b1-b8) in normalized magnitude obtained from the K2-C1 SC light curve of TW\,Vir. The black, red and blue light curves correspond to the type-I, II and III modulations shown in Figure~\ref{lctype}, respectively. The vertical dotted lines are the separators between the continuous two sections.}
\label{lc2}
\end{figure}

\begin{figure}
\centering
\includegraphics[width=14.0cm]{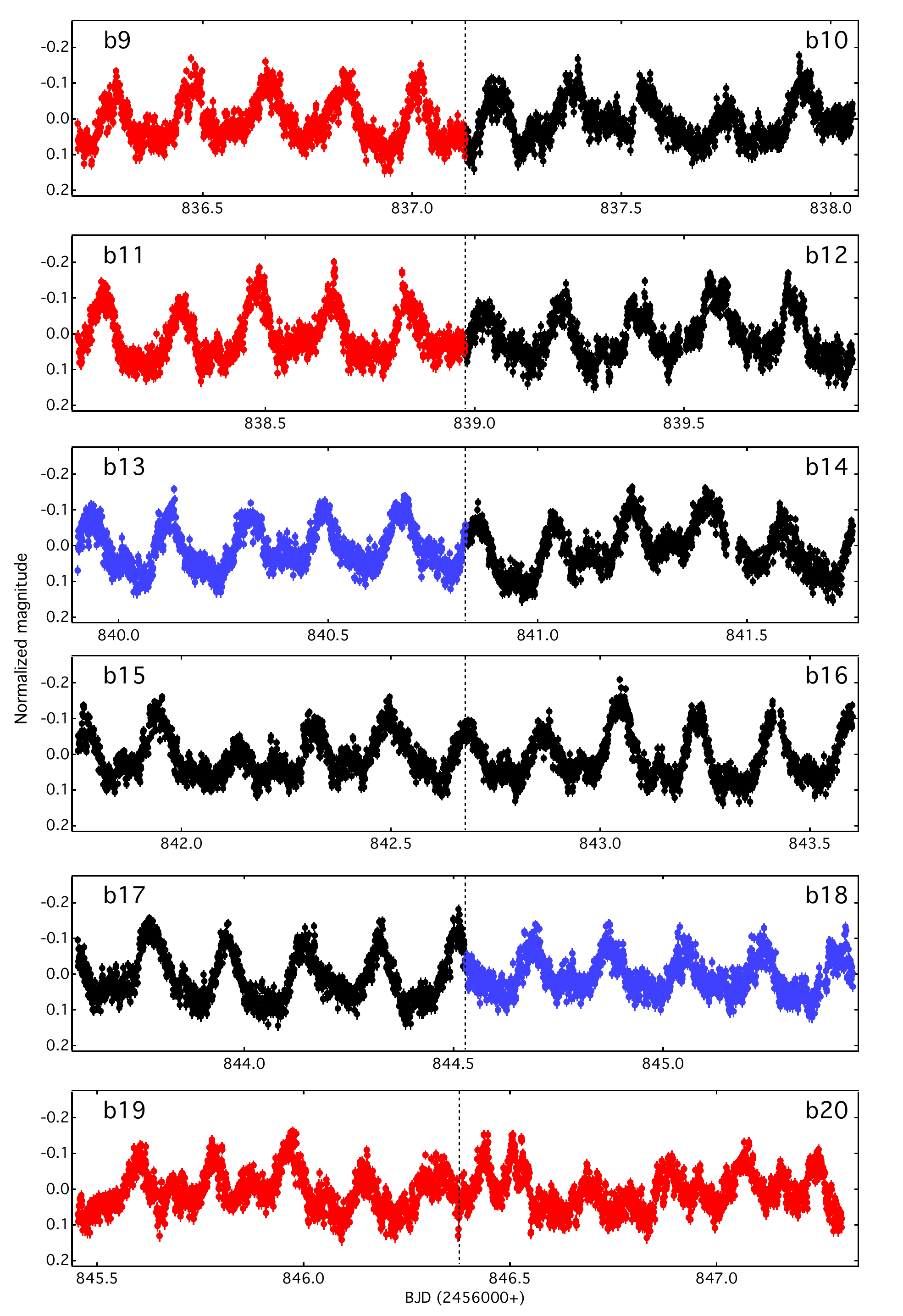}
\caption{The second half of the 24 sections (b9-b20). The colors of light curves are the same as those in Figure~\ref{lc2}.}
\label{lc3}
\end{figure}

Our data set started with the same 29 sections, including 5 before and 24 after the SO, as \citet{dai17} used to separate the quiescent K2-C1 SC light curve of TW\,Vir around the SO to improve the orbital period of TW\,Vir. In this paper, we carried out further analysis of these 29 sections with continuous start and end times. After several trials using the phase-correcting method described in \citet{dai17}, we empirically used the same data length of $\sim$\,0.93\,d covering five complete orbital cycles for each section to investigate the details of the disk variations around the SO. The trials showed if the length of each section is reduced by half ($\sim\,$0.47\,d), the cycle-to-cycle variations noticeably distort the orbital modulation, while if the length of each section is doubled ($\sim\,$1.86\,d), the insufficient number of sections smear the variations in orbital modulation. Hence, using shorter or longer sections does not provide needed details. Because large scatter in the first section before the SO overwhelms the regular modulation, and a stable double-hump modulation was not well established during the first 4 sections after the SO, these five sections were not considered in our synthetic analyses. The total light curve is shown in Figure~\ref{lc1}. The remaining 24 sections (4 before and 20 after the SO) are shown in Figures~\ref{lc2} and \ref{lc3}. An independent run of time-resolved power-spectra was done and agreed with Figure 7 of \citet{dai17}. Inspections of Figures~\ref{lc2} and \ref{lc3} indicate that all sections maintain the same luminosity state as that expected for the quiescent state of TW\,Vir. However, with the comprehensive K2 data, subtle variations in the light curves could be explored in this paper.

\begin{figure}
\centering
\includegraphics[width=16.0cm]{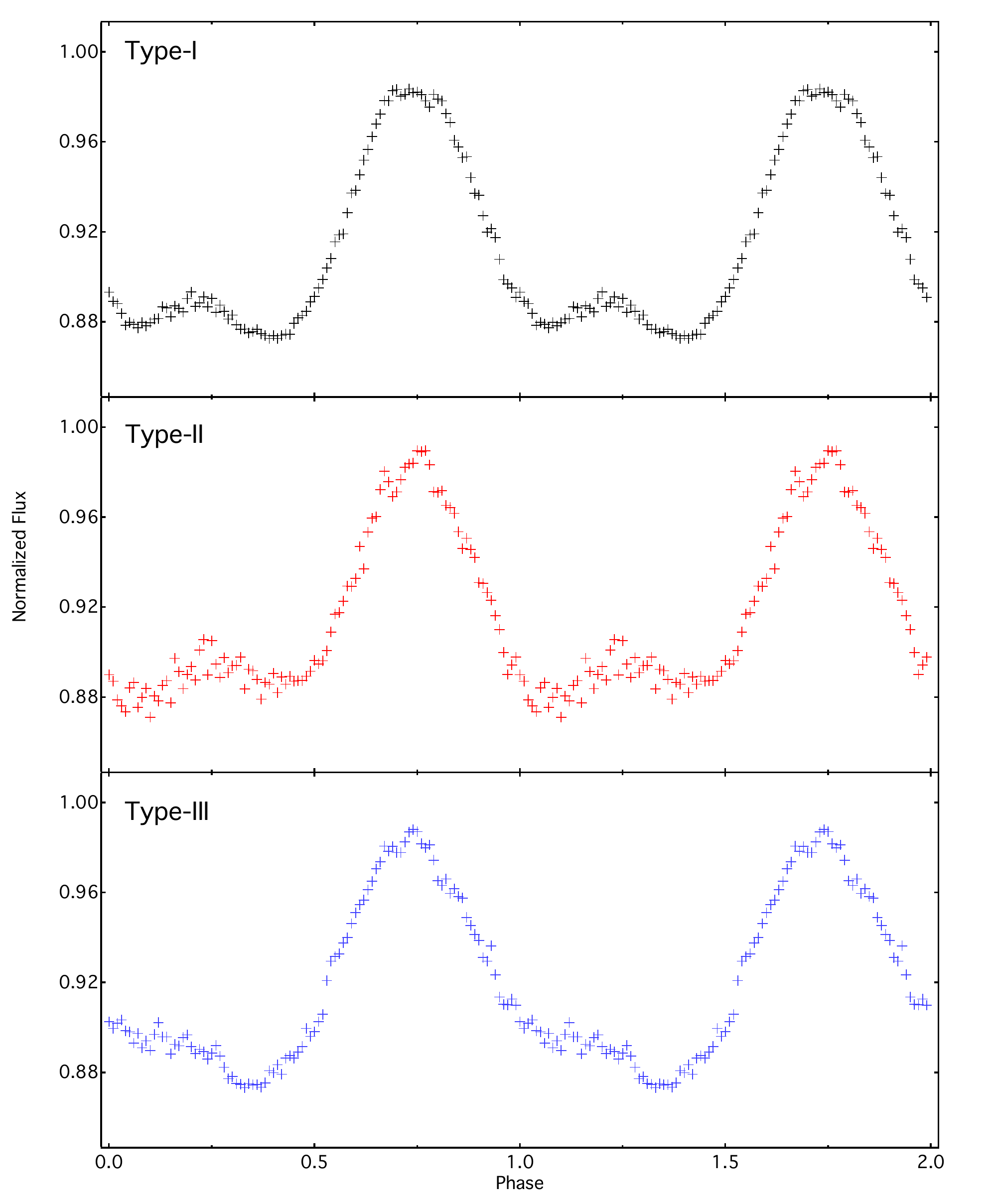}
\caption{From top to bottom, the black, red and blue phased light curves are the type-I, II and III orbital modulations in the normalized flux and folded on the orbital period of 0.182682(3)\,d. The line colors are the same as that in Figures~\ref{lc2} and~\ref{lc3}.}
\label{lctype}
\end{figure}

To provide the necessary format for an input file to XRBinary, each of the 24 sections was normalized to the maximum of the section and folded on the orbital period of 0.182682(3)\,d. By taking the derivative of the light curve, we obtained accurate phases for the two dips and humps present in each orbit by using where the first-order derivatives equal zero. Based on the different levels of the dips, the 24 sections were roughly separated into three types. The average light curve of ten sections with a higher-level minimum at phase zero (top panel of Figure~\ref{lctype}) is termed a type-I modulation (i.e., a typical double-hump light curve for DN). The middle panel of Figure~\ref{lctype} shows a type-II modulation averaged from six sections, where the minima levels are reversed and the secondary humps are "noisy". The remaining 8 sections shown in the bottom panel of Figure~\ref{lctype} illustrate a type-III modulation with a plateau rather than a typical secondary hump around phase 0.25. All three types of modulations commonly show a stable primary hump at phase 0.75 and a highly variable secondary hump around phase 0.25. Note that the changes in the three types of light curves are not in a successive order, but are likely related to stochastic processes caused by the white dwarf accretion, illustrating the variability in a typical CV during quiescence.

\section{Synthetic analysis}
\label{sec:sec3}

To further prepare the data for input to XRBinary, all 24 normalized and phased light curves were binned with a uniform phase resolution of 0.01. The model-2$^{\rm irra}$ and the disk height following a power law function with a fixed index of 1.1 are the same as that in \citet{dai18}. The parameters in the disk model of TW\,Vir derived by \citet{dai18} is preset to the initial parameters of the nonlinear fitting code NMfit. To focus on the disk variations before and after the SO, the six parameters q, i, $M_{\rm wd}$, $T_{\rm wd}$, $R_{\rm in}$ and $R_{\rm out}$\footnote{One difference between subtypes called U\,Gem- and SU\,UMa-type DN is whether a periodic modulation (i.e., superhump) accompanies a SO (those showing superhumps are generally SU\,UMa type). To explain the superhumps during a SO, the TTI model introduces a tidal 3:1 resonance instability between the disk and the secondary star to force a circular disk into a slowly precessing eccentric disk. For the U\,Gem-type DN TW\,Vir, $R_{\rm out}$=0.536(3)\,$R_{\odot}$ is smaller than a critical disk radius of $R_{\rm 3:1}\sim$0.67\,$R_{\odot}$ caused by the 3:1 Lindblad resonance tidal instability \citep{whi88,hir90,lub91}, and is $\sim$\,50\% of the white dwarf Roche lobe satisfying an empirical estimation for a quiescent DN \citep{har96}. Hence, a constant outer radius of the quiescent disk is a reasonable simplification for our synthetic light curve model, despite the detection of moderate variations in the disk radius before and after the SO in V1504\,Cyg and MLS\,0359+1750 \citep{osa13,lit18}.} are fixed during the calculations using NMfit, similar to the no-filter and V band disk models for KZ\,Gem in \citet{dai20}. Since the TTI model implies that a slowly precessing eccentric disk is not present at quiescence, the eccentricity of the disk is neglected in our disk model. The parameter uncertainties are estimated using the same method as that used in \citet{dai18}.

All 17 parameters consisting of 6 fixed and 11 adjustable parameters are listed in Table~\ref{tp}. The three adjustable parameters: $H_{\rm edge}$, $\mu$ and $L_{\rm d0}$ indicate that the radial effective temperature distribution of the disk changes during the 24 disk models corresponding to sections (a1-a4 and b1-b20). Including the two parameters $T_{\rm inner}$ (disk temperatures at $R_{\rm in}$) and $T_{\rm outer}$ (disk temperatures at $R_{\rm out}$), five parameters are used to describe the disk without hotspots. The edge-hotspot can be described by the three parameters: $T_{\rm es}$, $C_{\rm es}$ and $W_{\rm es}$. For the surface-hotspot, the two parameters $C_{\rm ss}$ and $W_{\rm ss}$, which are the centering phase and full width of the surface-hotspot in the angular direction, respectively, replace $\zeta_{\rm ssmin}$ and $\zeta_{\rm ssmax}$. In order to directly describe the temperature of the surface-hotspot, the parameter $T_{\rm ratio}$ is replaced with group parameters $T_{\rm ss0}$ (temperature of the disk surface surrounding surface-hotspot) and $T_{\rm ss}$ (temperature of surface-hotspot), similar to the other group parameters $R_{\rm ss0}$ and $R_{\rm ss}$ indicating the position of the surface-hotspot in the radial direction. Thus, a surface-hotspot needs six parameters: $C_{\rm ss}$, $W_{\rm ss}$, $T_{\rm ss0}$, $T_{\rm ss}$, $R_{\rm ss0}$ and $R_{\rm ss}$. With the luminosity of the two hotspots ($L_{\rm hs}$=$L_{\rm d}$-$L_{\rm d0}$, where $L_{\rm d}$ is the luminosity of disk with hotspots) and mass transfer rate from the secondary star to the disk ($\dot{M}_{\rm rd}\,\simeq\,L_{\rm acc}R_{\rm out}/GM_{\rm wd}$, where $L_{\rm acc}$ is the luminosity of the edge-hotspot), there are 16 parameters describing the disk model of TW\,Vir.

\begin{figure}
\centering
\includegraphics[width=12.0cm]{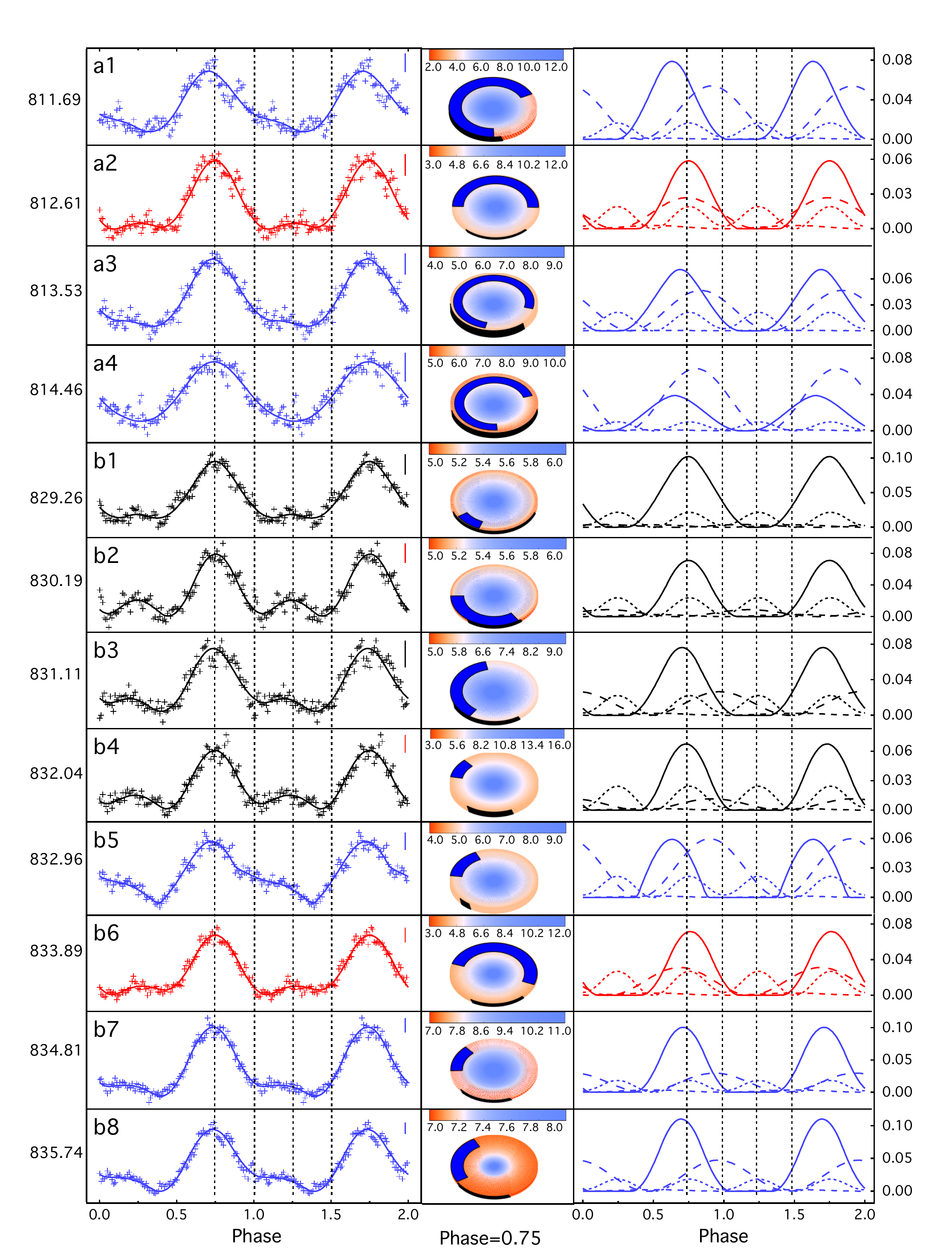}
\caption{The first half of the 24 disk models that correspond to the sections shown in Figure~\ref{lc2}. The phased and binned light curves superimposed with the best-fitting light curves are plotted in the left-hand panels. The small vertical solid lines plotted in the top right corner of all the left-hand panels denote the error bars of the binned light curves. The fluxes of all light curves are normalized. The median time of each section is labeled on the Y-axis of the left-hand panels. The colors of the light curves are the same as that in Figure~\ref{lc2}. All the middle panels show their corresponding 2D disk configurations at phase 0.75 using Phoebe 2.0. The colors in the 2D disk configuration denote the effective temperatures in units of 1000\,K. The secondary star is located on the right-hand side of the disk in a clockwise rotation direction. The relative flux contributions from four model components are plotted in the right-hand panels. The dotted and short dashed lines refer to the relative flux contributions from the two stellar components (white and red dwarfs) and the disk without the hotspots, respectively. The solid and long dashed lines denote the relative flux contributions from the edge-hotspot and the surface-hotspot, respectively, indicating which component is contributing to the actual light curve.}
\label{model1}
\end{figure}

\begin{figure}
\centering
\includegraphics[width=14.0cm]{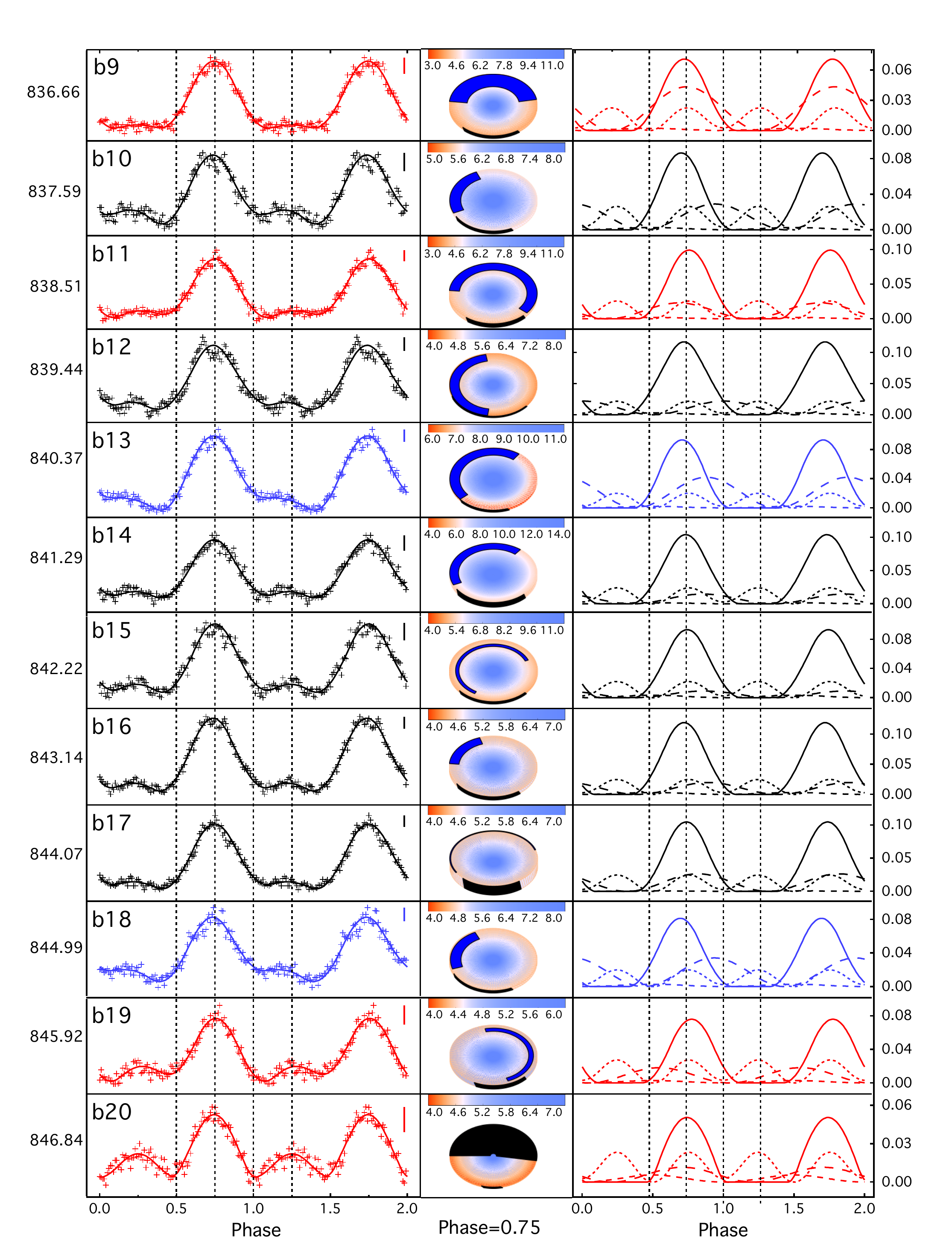}
\caption{Like Figure~\ref{model1}, the second half of the 24 disk models corresponding to the sections shown in Figure~\ref{lc3}. All symbols are the same as that of Figure~\ref{model1}. Note that a special surface-hotspot with $T_{\rm ss}$$<$\,$T_{\rm ss0}$ in the b20 disk model is filled with black rather than blue to distinguish this peculiarity.}
\label{model2}
\end{figure}

The secondary star in the typical 2D CV configuration (the middle panels of Figures~\ref{model1} and \ref{model2}) visualized by Phoebe 2.0\footnote{The version of Phoebe used for the CV plotting is 2.0a2 \citep{prs16}} can be omitted due to its identical size in all 24 disk models. To visualize the two hotspots on the disk, both are filled with black and dark blue rather than the color referred to in the color bar, because most of the temperature differences between the two hotspots and the surrounding regions of the disk are not large enough to clearly distinguish the hotspots from the disk. The black surface-hotspot shown in the bottom panel of Figure~\ref{model2} (b20) is an exception, denoting a lower temperature than that of the rest of the disk. The normalized and relative flux contributions from the four components: two component stars, the disk, and two hotspots are overlapped in the right-hand panels of Figures~\ref{model1} and~\ref{model2}. Their zero points\footnote{Since the zero points of the flux contributions from the edge-hotspot in all 24 disk models are equal to zero, they are replaced with the peaks of the flux contributions.} are plotted in Figure~\ref{rfc}. Despite the highest zero point of the relative flux contribution being from the disk, the disk only causes a small-amplitude sinusoidal modulation. The two component stars also only show a small-amplitude ellipsoidal modulation. Hence, the two hotspots dominate the morphology of the modulation. The edge-hotspot located near phase 0.75 explains the stable primary hump, while the surface-hotspot with its changing position gives rise to the variable secondary hump. Although the surface-hotspot changes very fast and without any apparent order, we found an interesting relationship between the peaks of the surface-hotspot contributions and the corresponding phases, as plotted in Figure~\ref{lctype2}. This diagram shows the three types of modulations separate into three different phase regions. Note that no peak contribution appears in the phase range of 0.2$\sim\,$0.5.

The relative measurements of the goodness of fit, $\chi^{2}$ (i.e., the variance between the calculated and observed light curves) for all 24 disk models shown in Figure~\ref{chi2} are in the range of 6$\sim$\,21. There are only two disk models (a1 and b12) with $\chi^{2}$$>$20, and the average $\chi^{2}$ is $\sim\,$10.7. 

\section{Quiescent Disk Evolution}
\label{sec:sec4}

Investigating the variations in all 16 parameters of the disk model of TW\,Vir shown in Figures~\ref{diskpars} and \ref{hotspotpars}, produces an evolution scenario of the quiescent disk around the SO. The SO separates the disk evolution into two stages in time order: pre-SO\footnote{In a strict sense, the four disk models (panels a1-a4 of Figure~\ref{model1}) just show the disk variations before the precursor rather than the SO. This precursor with a light maximum $\sim$\,0.5\,mag lower than the following SO is like the typical SU\,UMa-type DN (e.g., V1504\,Cyg and V344\,Lyr \citep{can10,can12,osa13,osa14}).} from quiescence to the SO, and post-SO to the next NO. The former stage covers $\sim\,$4\,d (a1-a4), far shorter than the latter stage with a coverage of $\sim\,$20\,d (b1-b20).

\subsection{Pre-SO}

\subsubsection{Disk}

Inspection of Figure~\ref{rfc} indicates that the relative flux contributions (from large to small) are the disk, the surface-hotspot, the two component stars and the edge-hotspot. In the last model (a4, $\sim$2 days prior to the peak of the precursor), the faint, cool disk suddenly switches to bright and hot. Assuming that $T_{\rm inner}$ and $T_{\rm es}$ are the indicators of the disk-white dwarf (mass accretion from the disk to the white dwarf occurring in the inner part of the disk) and stream-disk interactions (mass transfer from the secondary star to the disk occurring in the outer part of the disk), the decrease of $T_{\rm inner}$ and the small-amplitude variations in $T_{\rm es}$ from the a1 to a3 models indicate a surprise decrease of mass accretion but stable mass transfer. This implies that the disk continues to accumulate the material preparing for the upcoming SO supporting the DI model \citep{osa74,las01}, which requires a continuous accumulation of material to form a propagating and heating front on the disk. A stable $\dot{M}_{\rm tr}$ (panel e of Figure~\ref{diskpars}) further supports the DI model, rather than the EMT model\footnote{We cannot totally rule out a possibility that the enhanced mass transfer required by the EMT model appears during the following rising branch of the precursor.} \citep{sma91,sma04,sma08}.

\clearpage
\subsubsection{Hotspots}

The larger increment of the luminosity of the two hotspots ($L_{\rm hs}$) from the a3 to a4 model than that of $L_{\rm d0}$ suggests that the two hotspots produce the increase of the system light before the SO. Compared with the moderate edge-hotspot, the surface-hotspot located at phase $\sim\,$0.3, abruptly becomes much hotter and brighter in the a4 disk model (the beginning of the precursor), due to significant enhancement of $T_{\rm ss}$ and increase of the relative flux contribution from 25\% to $\sim\,$40\%. The contributions from the disk and the edge-hotspot reach $\sim\,$40\%, and $\sim\,$6\%, respectively, while the contribution from the two component stars in the a4 model suddenly drops to the lowest level of $\sim\,$9\%. The 2D CV configurations shown in the middle panels a1-a4 of Figure~\ref{model1} visually demonstrate that the surface-hotspot extends to almost comprise an annulus on the disk surface. Thus, we speculate that the surface-hotspot may be related to a heating front on the disk. When the surface-hotspot becomes hotter at the outer part of the disk, a heating front may be excited which then propagates the inner part of the disk inward to finally ignite the precursor and the following SO. The inward-propagating front causes the fast-rising light which is the signature of an outside-in outburst. This shape is similar to the two NO that follow the SO. All three outbursts being of an outside-in variety is consistent with the predictions of the TTI model \citep{osa89,osa96,osa05,osa13}.

\subsection{Post-SO}

\subsubsection{Disk}

In the first two models after the SO (b1 and b2), $T_{\rm inner}$ significantly decreases from $\sim$\,9400\,K (in the a4 model before the SO) to $\sim\,$5700\,K after the SO, then gradually increases up to a maximum of $\sim\,$16000\,K in the b4 model. From the b5 to b13 models, $T_{\rm inner}$ shows a small-amplitude fluctuation around an average of 9400\,K. After that, $T_{\rm inner}$ continuously decreases to $\sim\,$6200\,K similar to that in the b1 and b2 models. In contrast to the large-amplitude variations in $T_{\rm inner}$, $T_{\rm outer}$ is almost constant at $\sim\,$4000\,K with only a small-amplitude variation. Inspection of Figure~\ref{rfc} shows that the disk dominates the system light, similar to before the SO. A significant increase of the relative flux contribution from the disk after the SO implies that the disk that has brightened by the SO cannot immediately dim to its previous quiescent state before the SO.

\subsubsection{Hotspots}

\begin{figure}
\centering
\includegraphics[width=14.0cm]{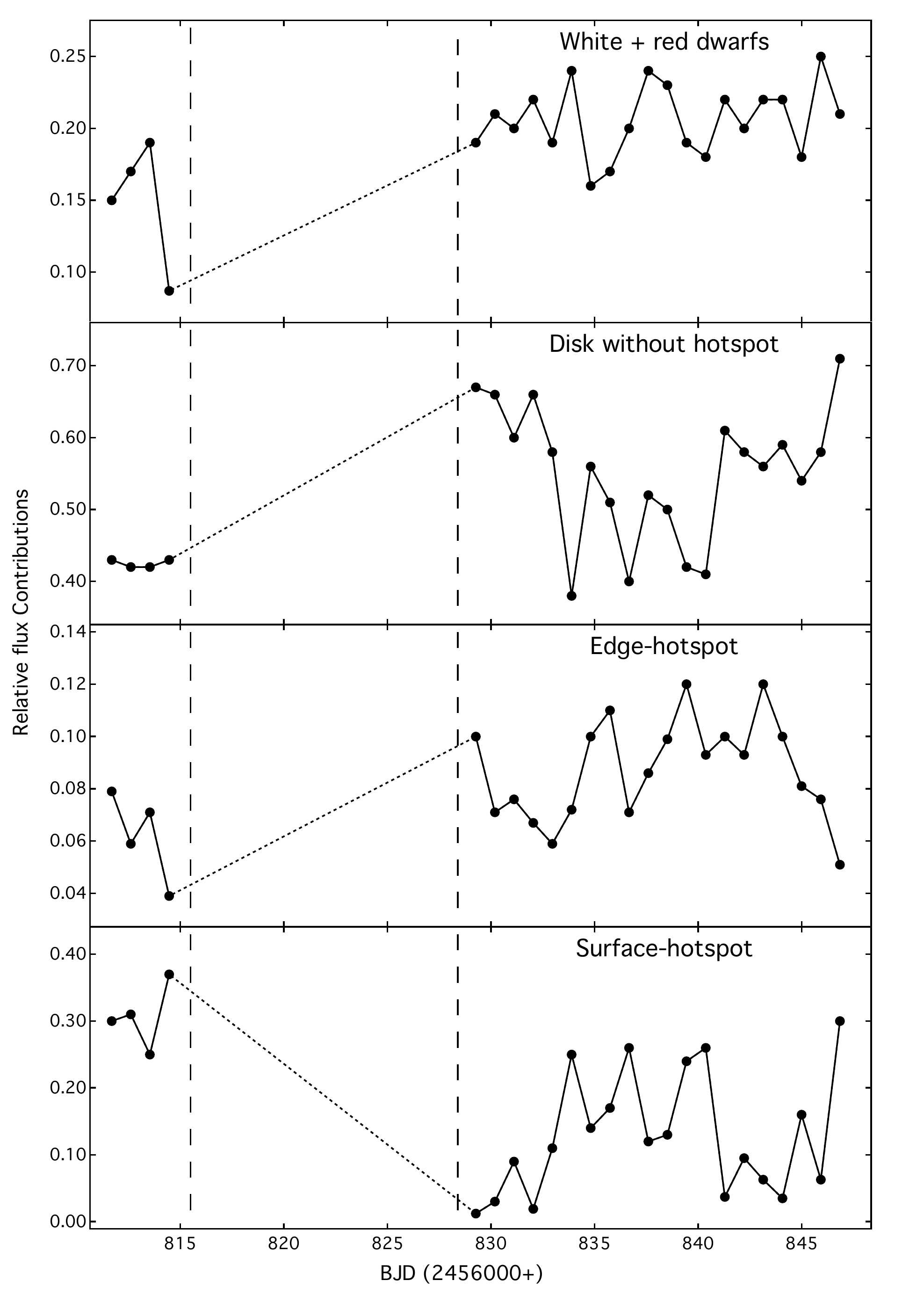}
\caption{From top to bottom, the variations in the zero points of the relative flux contributions from the four model components: the white+red dwarfs, the disk without hotspot, the edge-hotspot and the surface-hotspot, respectively. The two vertical dashed lines indicate the start and end of the SO.}
\label{rfc}
\end{figure}

\begin{figure}
\centering
\includegraphics[width=16.0cm]{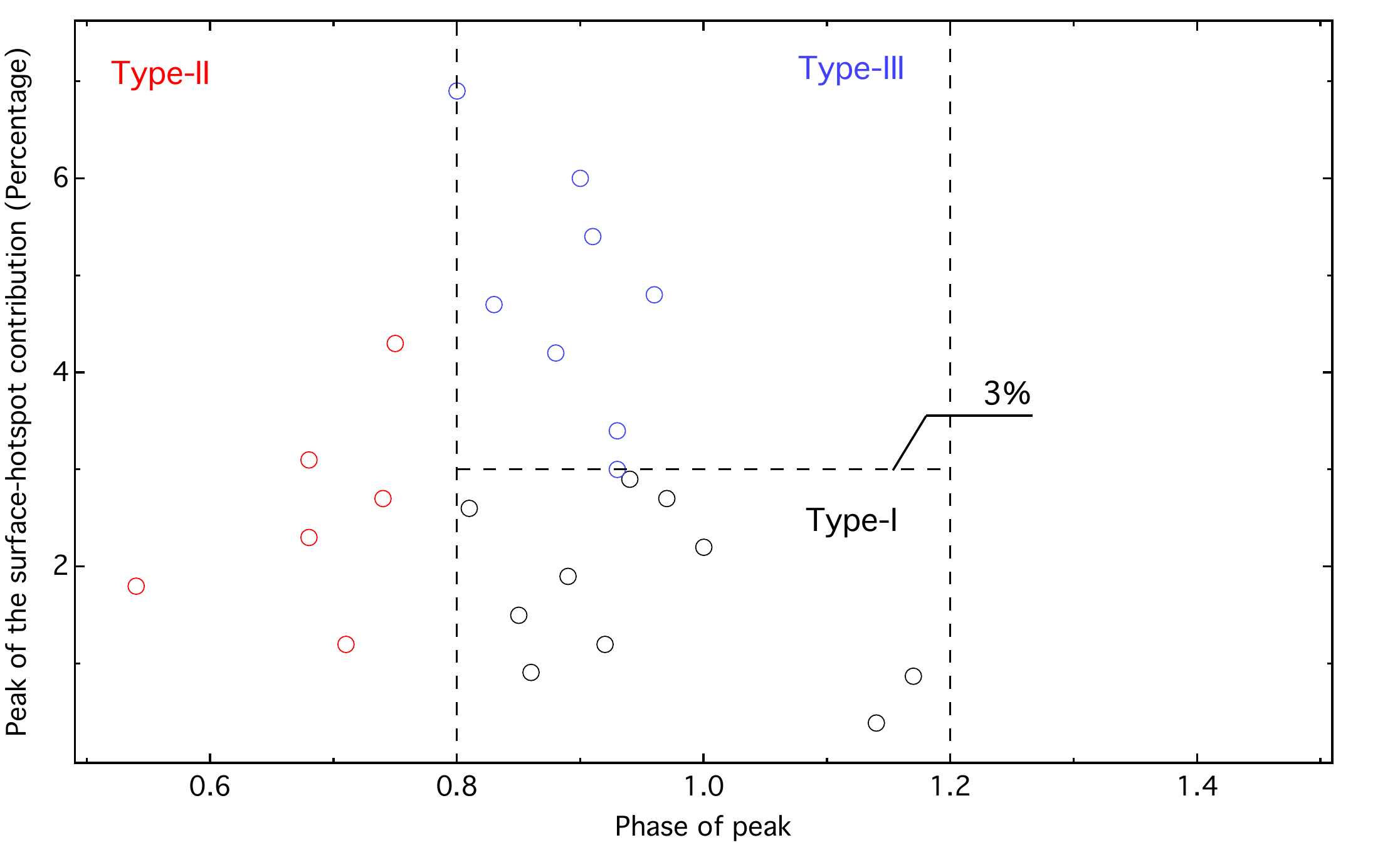}
\caption{The relationship between the peaks of the relative flux contributions from the surface-hotspot and the corresponding phases are plotted. The colors are the same as that in Figure~\ref{lctype}. The two vertical dash lines indicate phases 0.8 and 1.2, respectively. The horizontal dash line refers to the relative flux contribution of 3\%. All 24 sections are separated into three regions: the type-II modulations falling into the left region with a phase range of 0.5$\sim$\,0.8, the type-I and III modulations falling into the middle two regions lower and higher than 3\% with a phase range of 0.8$\sim$1.2. No data appears in the right region with a range of 1.2$\sim$\,1.5.}
\label{lctype2}
\end{figure}

\begin{figure}
\centering
\includegraphics[width=16.0cm]{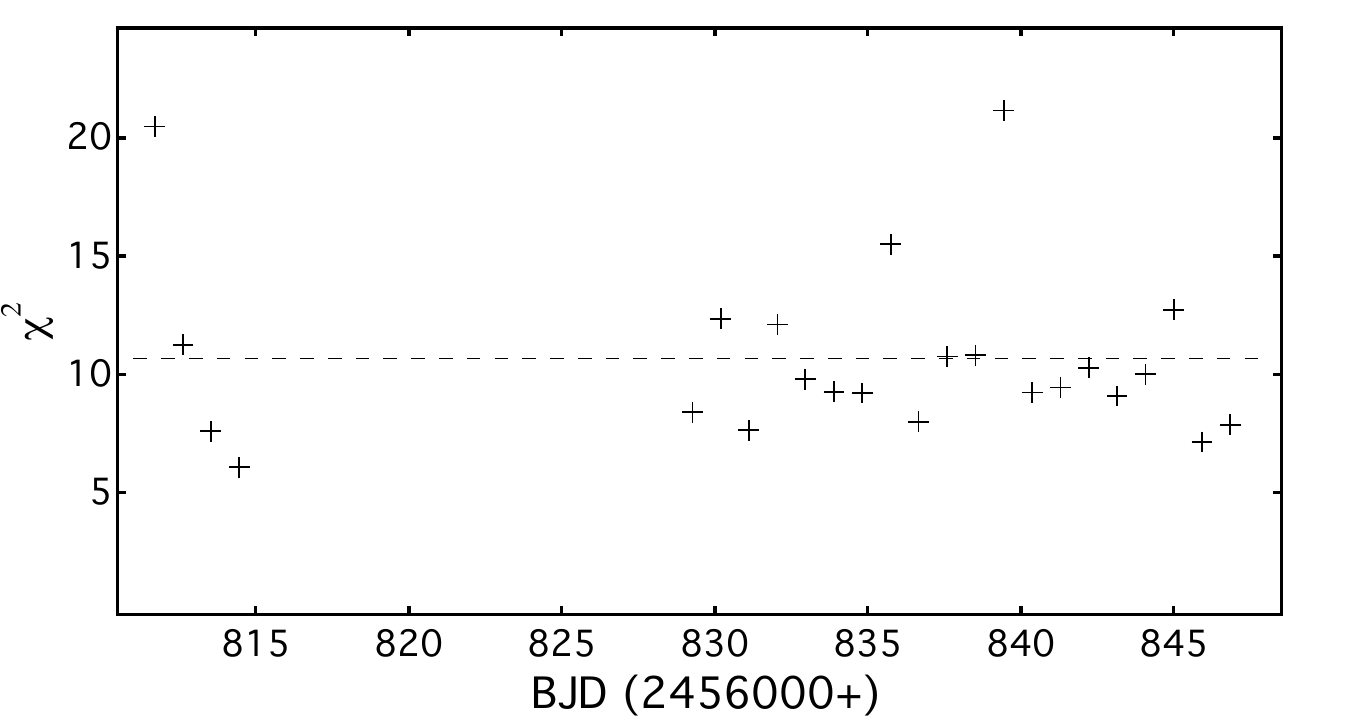}
\caption{Diagram of $\chi^{2}$ against time. The dashed line refers to the average.}
\label{chi2}
\end{figure}

\begin{figure}
\centering
\includegraphics[width=16.0cm]{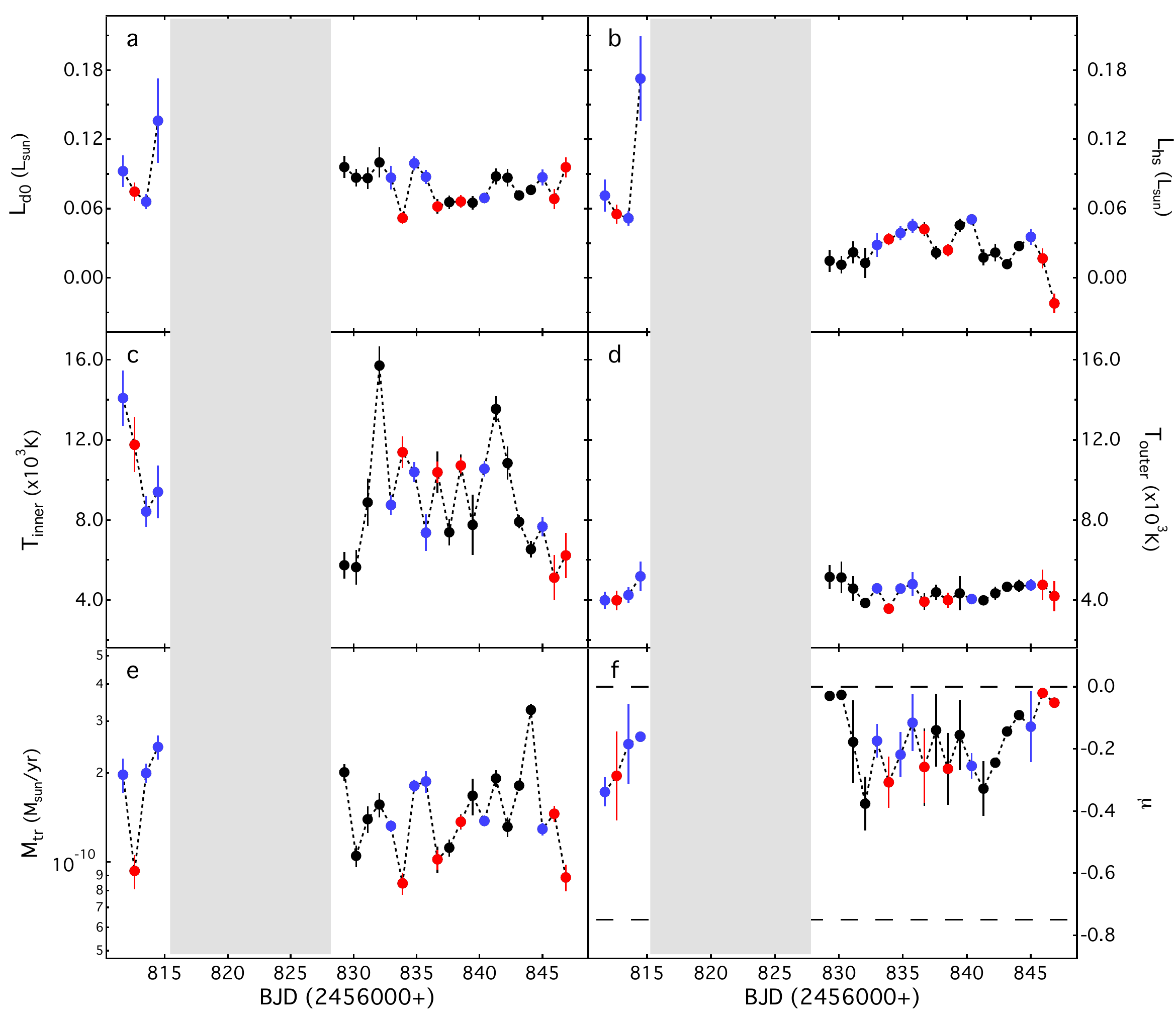}
\caption{The variations in 6 parameters of the disk model: $L_{\rm d0}$, $L_{\rm hs}$, $T_{\rm inner}$, $T_{\rm outer}$, $M_{\rm tr}$ and $\mu$ are plotted. The upper and lower lines in panel f correspond to $\mu$=-0.00 and -0.75, respectively. The light gray filled rectangle indicates a 15-days SO. The colors of the symbols are consistent with those of the light curves in Figures~\ref{lc2} and \ref{lc3}.}
\label{diskpars}
\end{figure}

\begin{figure}
\centering
\includegraphics[width=16.0cm]{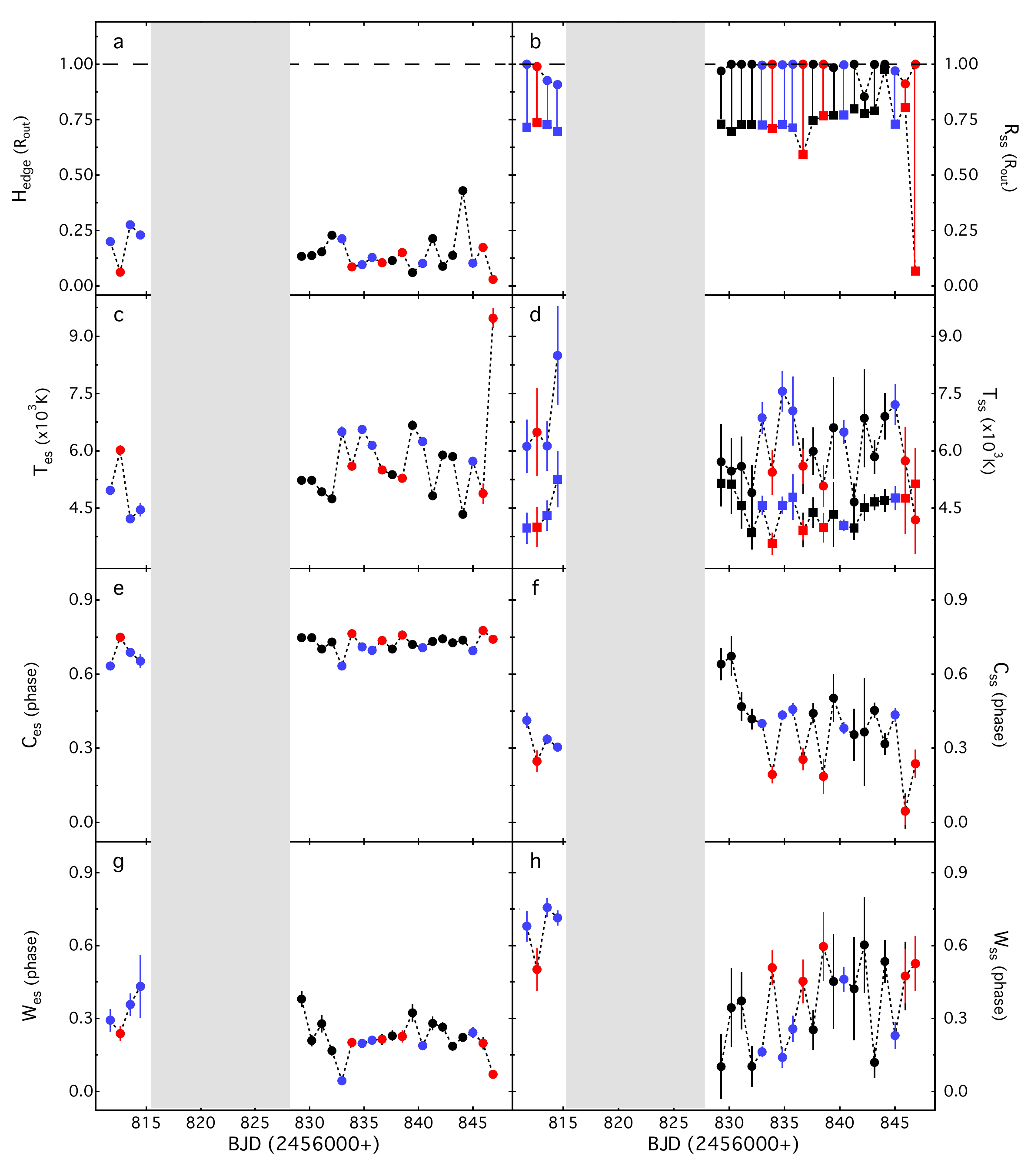}
\caption{The variations in two group parameters describing the edge-hotspot and the surface-hotspot are plotted in the left and right panels, respectively. For the surface-hotspot, $R_{\rm ss0}$ (solid squares) and $R_{\rm ss}$ (solid circles) are the two position parameters in the radial direction, and $T_{\rm ss0}$ (solid squares) and $T_{\rm ss}$ (solid circles) are the two temperature parameters as shown in two panels b and d, respectively. All symbols are the same as that of Figure~\ref{diskpars}.}
\label{hotspotpars}
\end{figure}

Panel b of Figure~\ref{diskpars} shows that the luminosity of the two hotspots significantly drops to a low, almost constant level indicating two stable hotspots after the SO. The edge-hotspot with a nearly constant peak in relative flux contribution ($\sim\,$8\% shown in Figure~\ref{rfc}) is present around phase 0.75 (the right panels of Figures~\ref{model1} and \ref{model2}). Inspection of panel g in Figure~\ref{hotspotpars} implies that the edge-hotspot becomes wider when approaching the SO, then has an average width of 0.2\,phase after the b5 model. The surface-hotspot located at the outer part of the disk ($R_{\rm ss0}$\,$>$\,0.77\,$R_{\rm out}$), with an almost constant width of 0.23\,$R_{\rm out}$\footnote{The abnormal b20 model is excluded and a small width of $<$0.08$R_{\rm out}$ only appears in the b15 and b17 models} in the radial direction. Assuming that this hotspot results from the release of the gravitational potential energy on the surface of the disk caused by the inward material flow through the disk \citep[cf.][]{can88,fra02,war03}, it could appear when the inward material flow crosses the low to high-density region. The nearly constant inner radius of the surface-hotspot ($R_{ss0}$) implies that this inward material flow encounters a stable transition region of disk density at 0.77\,$R_{\rm out}$ ($\sim\,$57\,$R_{\rm wd}$), causing an increase of the local temperature to finally form the hotspot. In spite of the constant radial position of the surface-hotspot shown in panel b of Figure~\ref{hotspotpars}, the surface-hotspot shows conspicuous differences before and after the SO, indicated by the variations in $T_{\rm ss}$, $C_{\rm ss}$, and $W_{\rm ss}$. The relative flux contribution from the surface-hotspot decreases from $\sim\,$40\% (a4) to $\sim\,$1\% (b1), the lowest contributor of the system light shown in Figure~\ref{rfc}. The two panels d and f in Figure~\ref{hotspotpars} show that during the b1-b4 models, the surface-hotspot with a low average temperature of 5500\,K, gradually shifts from phase 0.7 to 0.4, close to that before the SO. Between the b5 and b18 models, the surface-hotspot with a higher temperature remains at phase $\sim\,$0.4.

\section{Discussion}
\label{sec:sec5}

\subsection{Hotspots before the NO}

The last disk model (b20) before the NO shows two hotspots on a normal disk with a relative flux contribution increasing up to $\sim\,$70\%. The edge-hotspot located at phase 0.75 similar to that in the previous models, shows the highest temperature of $T_{\rm es}$$\sim$$10^{4}$\,K and the smallest width in the angular direction of $W_{\rm es}$$\sim$0.07\,phase. Assuming that the enhanced $T_{\rm es}$ is caused by the increased mass transfer rate from the secondary star, the disappearance of a stable orbital modulation in the following NO interval may be explained by unstable mass transfer\footnote{Although this assumption seems to support the MTI model \citep{sma91}, the K2-C1 data gap after BJD\,2456846.84 prevents any further investigation of the following variations in the disk and hotspots.}.

In opposition to the edge-hotspot, the surface-hotspot, with a similar position and width in the angular direction, shows the largest coverage of 0.93\,$R_{\rm out}$ in the radial direction (i.e., $R_{\rm ss0}$ is suddenly lowered from an average of $\sim\,$57\,$R_{\rm wd}$ to $\sim\,$5\,$R_{\rm wd}$ close to the white dwarf surface), and the lowest temperature of $T_{\rm ss}$$\sim$\,4000\,K obviously below $T_{\rm ss0}$$\sim\,$5000\,K. Thus, the surface-hotspot is actually a ``coolspot"\footnote{To recheck this abnormal surface-hotspot, we carried out several trials with different initial parameter sets to search for other possible results with a normal $T_{\rm ss}$. Although some convergent results were obtained, the best-fitting model shows the lowest $\chi^{2}$$\sim$7.8.}, visualized in panel b20 of Figure~\ref{model2}. Assuming that the other half of the disk with the higher temperature is regarded as a surface-hotspot (i.e., the ``coolspot" is opposite its normal position on the disk), this anomalous surface-hotspot could imply that the disk is on the way to switching into a hot status (i.e., the following NO).

\subsection{Quiescent disk temperature distribution}

\begin{figure}
\centering
\includegraphics[width=16.0cm]{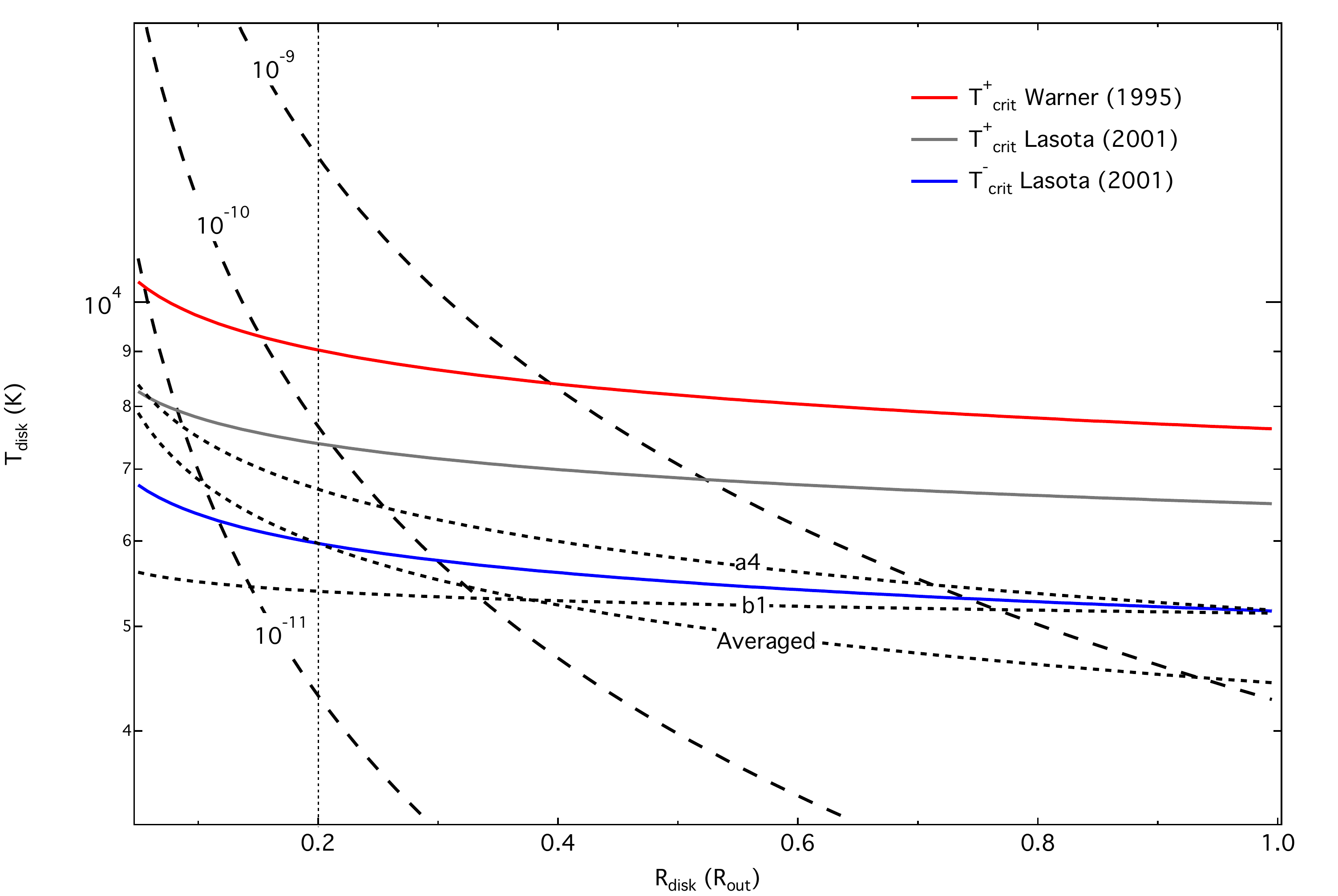}
\caption{The short-dashed lines denote the radial effective temperature distributions of the averaged, a4 and b1 disk models. Three long-dashed lines refer to the radial effective temperature distributions of a steady-state optically thick disk under three constant mass transfer rates: $10^{-9}$, $10^{-10}$ and $10^{-11}$ in units of $M_{\odot}$\,yr$^{-1}$ marked by the digits superimposed on the lines. The outer disk region located at the right side of the vertical dotted line at $R_{\rm disk}$$\sim$\,0.2\,$R_{\rm out}$ shows everywhere $T_{\rm disk}<\,T_{\rm crit}^{-}$ in quiescence.}
\label{dtd}
\end{figure}

Based on the energy dissipation rate in the disk, \citet{fra02} proposed that the effective temperature of a steady-state optically thick disk is a power-law function of the disk radius ($T_{\rm disk}$$\propto$$R_{\rm disk}^{-3/4}$) under a constant mass transfer rate from the secondary star. Inspection of panel f of Figure~\ref{diskpars} indicates that the quiescent disk of TW\,Vir is not a steady-state thick disk due to the derived flat radial temperature distribution (Figure~\ref{dtd}). Along with the variations in $T_{\rm inner}$, the power index shows an opposite variation in that $\mu$ continuously increases up to near zero before the SO (a1-a4), and decreases from near zero (b1 and b2) to a minimum of -0.38 (b4), then oscillates around an average of -0.22 (b5-b13), and finally recovers to be near zero again at the end of the quiescence before the NO. This implies that the disk temperature distribution of TW\,Vir is much flatter when approaching outburst, like the short-period DN V4140\,Cyg with $\mu$\,$\simeq$\,-0.36 in quiescence and $\mu$\,$\simeq$\,-0.25 in outburst \citep{bor05}. In contrast, the radial temperature distributions of Z\,Cha and OY\,Car are flat in quiescence with $\mu$$>$-0.2 and almost follow the power-law function of a steady-state optically thick disk in outburst with $\mu$$<$-0.7 \citep{hor85,woo90,rut92}. \citet{ida10} further claimed that all observed DN in quiescence have flat radial-temperature profiles in excellent agreement with the prediction of the DI model.

According to the standard DI model, the two critical effective temperatures, $T_{\rm crit}^{+}$ and $T_{\rm crit}^{-}$, corresponding to the principal critical values of the minimal and maximal surface-densities of the disk, $\Sigma_{\rm min/max}$, can be estimated by the following two formulas \citep{las01},
\begin{equation}
T_{\rm crit}^{+}\,=\,7200\,\alpha_{\rm H}^{-0.002}\,(\frac{M_{\rm wd}}{M_{\odot}})^{0.03}(\frac{R_{\rm disk}}{10^{10}\rm cm})^{-0.08}\quad\rm Kelvin,
\end{equation}
\begin{equation}
T_{\rm crit}^{-}\,=\,5800\,\alpha_{\rm C}^{-0.001}\,(\frac{M_{\rm wd}}{M_{\odot}})^{0.03}(\frac{R_{\rm disk}}{10^{10}\rm cm})^{-0.09}\quad\rm Kelvin,
\end{equation}
where $M_{\rm wd}$=1.1\,$M_{\odot}$ for TW\,Vir \citep{dai18}, $\alpha_{\rm H}$ and $\alpha_{\rm C}$ are the two constant disk viscosities \citep{sha73} on the upper (hot) and lower (cold) branches of the S-curve (i.e., an S-shaped curve shown in the $\Sigma$-$T_{\rm eff}$ plane), respectively. In spite of the complicated descriptions of $\alpha$ \citep{hub98,kro07}, $T_{\rm crit}^{+}$ and $T_{\rm crit}^{-}$ are almost independent of $\alpha$ since the power-index of $\alpha$ is close to zero. The $T_{\rm crit}^{+}$, $T_{\rm crit}^{-}$ and the averaged $T_{\rm disk}$ of the 24 disk models are shown in Figure~\ref{dtd}. The a4 disk model shows that the disk is at a temperature between $T_{\rm crit}^{-}$ and $T_{\rm crit}^{+}$, indicating the start of disk instability, while $T_{\rm disk}$ of the b1 disk model is all below $T_{\rm crit}^{-}$. The averaged $T_{\rm disk}$ in the inner part of a disk ($<$\,0.2$R_{\rm out}$$\simeq$15\,$R_{\rm wd}$) cannot satisfy the DI requirement of $T_{\rm disk}<\,T_{\rm crit}^{-}$ in quiescence. This seems to be in contradiction with the DI model, similar to V4140\,Sgr \citep{bap16}. Hence, a simple, uniform power-law function may not be enough to describe the quiescent disk temperature distribution of a CV disk, especially in the inner part of the disk. We speculate that this discrepancy is caused by our preset disk extending down to the white dwarf surface, while the inner part of the disk may be truncated by evaporation of the inner disk \citep[cf.][]{ham99,men00,dub01}. Note that using a formula from \citet{war95}, $T_{\rm crit}^{+}$=7690($R_{\rm disk}$/3$\times$10$^{10}$ cm)$^{-0.105}$M$_{\rm wd}^{0.15}$ Kelvin, most of the 24 quiescent disk models show $T_{\rm disk}$ below this $T_{\rm crit}^{+}$. This is roughly consistent with the loose constraint of the DI model, $T_{\rm disk}$$<$\,$T_{\rm crit}^{+}$.

\subsection{Mass transfer rate at quiescence}

Despite the large-amplitude variations in the temperature of the edge-hotspot (excluding the special b20 model) at quiescence shown in panel c of Figure~\ref{hotspotpars}, the mass transfer rate, $\dot{M}_{\rm tr}$, estimated from the edge-hotspot (panel e of Figure~\ref{diskpars}) shows a small-amplitude variation around an average of 1.57$\times10^{-10}$\,$M_{\odot}$\,yr$^{-1}$, and a trivial maximum of $\sim\,$3.3$\times10^{-10}$\,$M_{\odot}$\,yr$^{-1}$ in the b17 model. We found that this increased $\dot{M}_{\rm tr}$ does not correspond to an enhanced edge-hotspot, but to a thickened disk with a maximum of $H_{\rm edge}$$\sim\,$0.43\,$R_{\rm out}$.

\begin{figure}
\centering
\includegraphics[width=16.0cm]{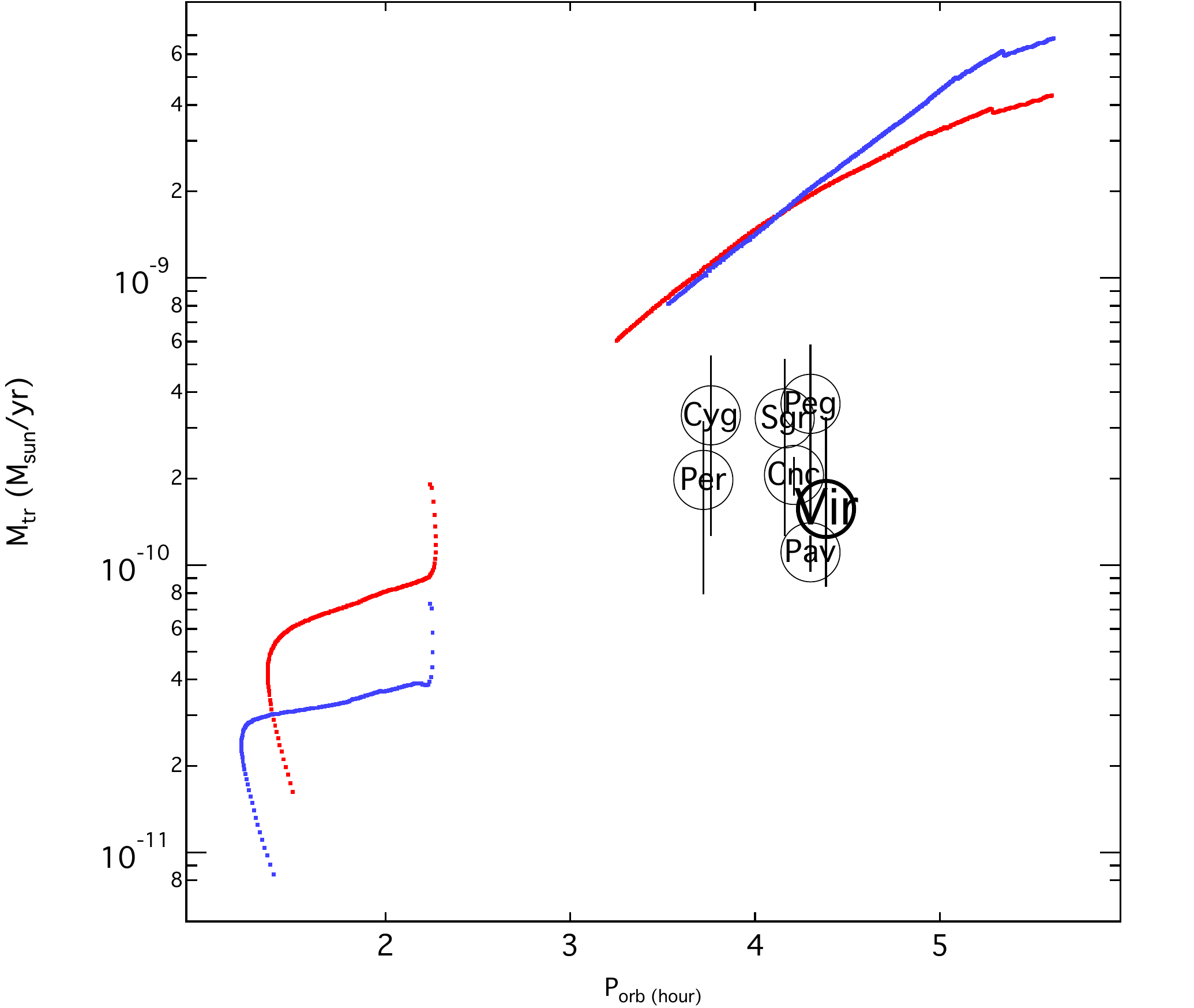}
\caption{Mass transfer rates as a function of the orbital period. The red and blue lines refer to the mass transfer rates expected by the standard and revised models \citep{kni11}, respectively. The constellation names represent the CVs listed in Table~\ref{cvname}.}
\label{dmdt}
\end{figure}

To compare with the expected secular mass transfer rate at the orbital period of TW\,Vir \citep{kni11}, $\dot{M}_{\rm tr}$ in all 24 disk models are averaged, and the maximum and minimum of $\dot{M}_{\rm tr}$ were set to the upper and lower limits of the error bar. This averaged $\dot{M}_{\rm tr}$ plotted in Figure~\ref{dmdt} is significantly lower than the corresponding secular mass transfer rate derived by the standard/revised models. In spite of this large deviation, many other U\,Gem-type DN with orbital periods in the range of 3-5\,hr listed in Table~\ref{cvname} \citep{dub18} have a similar mass transfer rate to TW\,Vir. The mass transfer rates of these U\,Gem-type DN appear to be overestimated by the standard/revised models.

\section{Conclusions}
\label{sec:sec6}

The quiescent K2-C1 SC light curve of TW\,Vir separated into 24 sections with an optimal data length of 0.93\,d, including 4 before and 20 after a SO, shows morphological differences that can be roughly classified into three types based on different levels of the light minima. Changes in the orbital modulation appear randomly but a stable primary hump caused by an edge-hotspot is present at phase 0.75 while a lower-amplitude secondary hump is highly variable. The differences in these three types are plausibly explained by a surface-hotspot with different positions and intensities.

Using the non-linear fitting code XRBinary and NMfit, 24 disk models for the corresponding 24 sections are calculated. The model parameters do not show large differences before and after the SO, demonstrating that the accretion pattern is not broken by the SO. All 24 disk models indicate that the disk dominates the flux contributions to the quiescent system light. Based on the variations of the 16 parameters in the 24 disk models, a complete quiescent disk evolution scenario around the SO is obtained. Below is a summary of our findings:

\begin{itemize}

\item{Pre-SO: The mass accretion from the disk to the white dwarf gradually declines, while the mass transfer from the secondary to the disk remains stable before a precursor to the SO occurs.}

\item{Precursor: The start of the precursor shows an enhanced and ringlike surface-hotspot located at phase $\sim\,$0.3. The faint and cool disk switches to be bright and hot. A stable edge-hotspot develops during this stage.}

\item{Post-SO: The outer part of the disk remains almost at a constant temperature of $\sim$\,4000\,K similar to that in the pre-SO stage. However, the inner part of the disk experiences a large variation starting with a cool state at the end of the SO. The edge-hotspot with an average angular width of 0.2 phase is located at phase 0.75. The surface-hotspot with a large-amplitude variation in temperature remains almost constant (0.77-1.0\,$R_{\rm out}$) in the radial direction, while showing large changes in the angular direction.}

\item{End of quiescence: Before the first NO following the SO, the edge-hotspot abruptly becomes much hotter, and, the surface-hotspot changes to an anomalous ``coolspot" covering over half of the disk surface. The derived radial temperature distribution of the disk is flat at quiescence with a disk power-law $>$\,-0.38 and much flatter when approaching the outbursts.}

\item{A mass transfer rate estimated from the edge-hotspot is 0.8-3.3$\times$$10^{-10}$\,$M_{\odot}$\,yr$^{-1}$, typical for U\,Gem-type DN with orbital periods of 3-5\,hr, but far lower than the predictions of standard/revised models.}

\end{itemize}
 
\acknowledgments

This work was partly supported by CAS Light of West China Program, the Chinese Natural Science Foundation (No. 11933008), and the Science Foundation of Yunnan Province (No. 2016FB007). PS acknowledges support from NSF grant AST-1514737. We thank Colin Littlefield for his time-resolved power-spectra, and Edward L. Robinson for his helpful suggestions on XRBinary program.

\software{XRBinary (v2.4), NMfit (v2.0), Phoebe \citep[v2.0;][]{prs16})}

\begin{table}
\caption{The initial parameters of TW\,Vir used in the synthetic analysis.}
\begin{center}
\label{tp}
\begin{tabular}{llcc}
\hline\hline
& Parameters & \tablenotemark{a}Values & Statements\\
\hline
\textbf{Fixed} &&&\\
\hline
& q($M_{\rm rd}/M_{\rm wd}$) & 0.41(5) & mass ratio\\
& i(degree) & 44.3(5) & orbital inclination\\
& $M_{\rm wd}$($M_{\odot}$) & 1.10(3) & white dwarf mass\\
& $T_{\rm rd}$($\times10^{3}\,$K) & 4.00(4) & red dwarf temperature\\
& \tablenotemark{b}$R_{\rm in}$($R_{\odot}$) & 0.007 & inner radius of the disk\\
& $R_{\rm out}$($R_{\odot}$) & 0.536(2) & outer radius of the disk\\
\hline
\textbf{Adjustable} &&&\\
\hline
& $H_{\rm edge}$($R_{\odot}$) & 0.1174 & height of the disk at the outer edge\\
& \quad$\mu$ & -0.11 & power index of the disk temperature\\
& $L_{\rm d0}$($\times10^{32}\rm{erg\,s}^{-1}$) & 3.36 & luminosity of the disk without hotspots\\
& $T_{\rm es}$($\times10^{3}\,$K) & 4.80 & temperature of edge-hotspot\\
& $C_{\rm es}$(phase) & 0.727 & centering phase of edge-hotspot\\
& $W_{\rm es}$(phase) & 0.296 & full width of edge-hotspot\\
& \multirow{2}*{$R_{\rm ss0}$($R_{\odot}$)} & \multirow{2}*{0.489} & lower limit of boundary of surface-hotspot\\
&&&in the radial direction\\
& \multirow{2}*{$R_{\rm ss}$($R_{\odot}$)} & \multirow{2}*{0.536} & upper limit of boundary of surface-hotspot\\
&&&in the radial direction\\
&  \multirow{2}*{$\zeta_{\rm ssmin}$(phase)} &  \multirow{2}*{0.32} & lower limit of boundary of surface-hotspot\\
&&&in the radial direction\\
& \multirow{2}*{$\zeta_{\rm ssmax}$(phase)} & \multirow{2}*{0.74} & upper limit of boundary of surface-hotspot\\
&&&in the radial direction\\
& \quad$T_{\rm ratio}$ & 1.20 & fractional change in the temperature of surface-hotspot\\
\hline\hline
\end{tabular}
\end{center}
\textbf{\quad Note.}
\tablenotetext{\quad a}{\quad\, The values of parameters and uncertainties are calculated from an average light curve after the SO \citep{dai18}.}
\tablenotetext{\quad b}{\quad\, $R_{\rm in}$ is preset to be equal to the white dwarf radius.}
\end{table}

\begin{table}
\caption{Mass transfer rates of the DN with orbital periods in a range of 3-5\,hr.}
\begin{center}
\label{cvname}
\begin{tabular}{lcl}
\hline\hline
\tablenotemark{a}CV name & \tablenotemark{b}Subtype & \tablenotemark{c}$\dot{M}_{\rm tr}$\\
\hline
BD Pav & UG & $1.1(\pm0.2)$\\
GY Cnc & UG & $2.1(\pm0.3)$\\
PY Per & UGZ & $2.0(\pm1.2)$\\
V729 Sgr & UG & $3.2(\pm2.0)$\\
V513 Peg & UG & $3.6(\pm2.2)$\\
V811 Cyg & UGSS & $3.3(\pm2.1)$\\
\tablenotemark{d}TW Vir & UG & $1.6(^{+1.7}_{-0.7})$\\
\hline\hline
\end{tabular}
\end{center}
\textbf{\quad Note.}
\tablenotetext{\quad a}{\quad\, The constellation names are marked in Figure~\ref{dmdt}.}
\tablenotetext{\quad b}{\quad\, Abbreviations are the same as Table A.2 in \citet{dub18}.}
\tablenotetext{\quad c}{\quad\, The unit of $\dot{M}_{\rm tr}^{c}$ is $10^{-10}M_{\odot}$\,yr$^{-1}$.}
\tablenotetext{\quad d}{\quad\, calculated in this paper.}
\end{table}

\end{CJK*}
\end{document}